\documentclass[a4paper,11pt]{article}

\usepackage{amsfonts, amsmath, amssymb}
\usepackage[utf8]{inputenc}
\usepackage[a4paper,top=3cm,bottom=2.5cm,left=2.5cm,right=2.5cm, bindingoffset=5mm]{geometry}
\usepackage{color}
\usepackage{booktabs}
\usepackage{subfigure}
\usepackage{amssymb}
\usepackage{amsmath}
\usepackage{amsfonts}
\usepackage{bm}
\usepackage{pifont}
\usepackage{graphicx}% Include figure files
\usepackage{float}
\usepackage{hyperref}
\usepackage{multirow}
\hypersetup{citecolor=blue}
\usepackage{array}
\usepackage{multirow}
\usepackage{colordvi}

\newcommand{\be}{\begin{equation}}
\newcommand{\ee}{\end{equation}}
\newcommand{\beq}{\begin{equation*}}
\newcommand{\eeq}{\end{equation*}}
\newcommand{\ba}{\begin{eqnarray}}
\newcommand{\ea}{\end{eqnarray}}

\def\bea{\begin{eqnarray}}
\def\eea{\end{eqnarray}}

\newcommand{\ov}{\overline}

\newcommand{\bra}[1]{\ensuremath{\langle #1 |}}   % Bra vector
\newcommand{\ket}[1]{\ensuremath{| #1 \rangle}}   % Ket vector

\newcommand{\eps}{\varepsilon}
\newcommand{\ldm}{\ensuremath{{\Delta m_{31}^2}}}          % Mass squared differences
\newcommand{\sdm}{\ensuremath{{\Delta m_{21}^2}}}

\begin{document}

\vspace*{-23mm}
\begin{flushright}
SISSA 23/2014/FISI\\
RM3-TH/14-7
\end{flushright}
\vspace*{0.5cm}

\vspace{20pt}
\begin{center}
{\bf{\large  The Daya Bay and T2K results on $\sin^2 2 \theta_{13}$ 
and Non-Standard Neutrino Interactions}}

\vspace{0.4cm} I. Girardi$\mbox{}^{a)}$, D. Meloni$\mbox{}^{b)}$ and S. T. Petcov$\mbox{}^{a,c)}$
\footnote{Also at: Institute of Nuclear Research and Nuclear Energy,
Bulgarian Academy of Sciences, 1784 Sofia, Bulgaria.}

\vspace{0.4cm} $\mbox{}^{a)}${\em  SISSA/INFN, Via Bonomea 265, 34136
Trieste, Italy.\\}

\vspace{0.1cm} $\mbox{}^{b)}${\em  \it  Dipartimento di Matematica e Fisica, 
Universit\`{a} di Roma Tre,}
\centerline{\it  Via della Vasca Navale 84, I-00146 Rome.}

\vspace{0.1cm} $\mbox{}^{c)}${\em Kavli IPMU (WPI), The University
of Tokyo, Kashiwa,
Japan.\\
}

\end{center}

\begin{abstract}
We show that the relatively large best fit value of 
$\sin^2 2 \theta_{13} = 0.14 \, (0.17)$
measured in the T2K experiment for fixed values of 
i) the Dirac CP violation phase $\delta = 0$, 
and ii) the atmospheric neutrino mixing parameters
$\theta_{23} = \pi/4$,  $|\Delta m^2_{32}| = 
2.4 \times 10^{-3} \; {\rm eV}^2$,
can be reconciled with the Daya Bay result 
$\sin^2 2 \theta_{13} = 0.090 \pm 0.009$ 
if the effects of non-standard neutrino interactions (NSI) 
in the relevant $\bar \nu_e \to \bar \nu_e$ 
and $\nu_\mu \to \nu_e$ oscillation probabilities 
are taken into account.
\end{abstract}

\section{Introduction}

Recently the T2K collaboration reported a measurement of
the reactor neutrino mixing angle $\theta_{13}$ based on their 
latest $\nu_{\mu} \rightarrow \nu_e$ oscillation data \cite{Abe:2013hdq}.
Fixing the values of i) the Dirac CP violation (CPV) phase $\delta = 0$,
ii) the atmospheric neutrino mixing angle $\theta_{23} = \pi/4$,
iii) $\sin^2 \theta_{12} = 0.306$, iv) $\sdm = 7.6 \times 10^{-5} \; {\rm eV}^2$
and v) $|\Delta m^2_{32} | = 2.4 \times 10^{-3} \; {\rm eV}^2$,
the T2K collaboration found:
\be
\sin^2 2\theta_{13} = 0.140^{+0.038}_{-0.032} \, (0.170^{+0.045}_{-0.037}) \;,
\label{eq:T2Kth13}
\ee 
where the value (the value in brackets) corresponds to neutrino mass spectrum
with normal (inverted, IO) ordering (NO). The best fit value of $\sin^2 2 \theta_{13}$
reported by the T2K collaboration is significantly larger than
that measured in the reactor neutrino experiments Daya Bay, RENO and Double Chooz
\cite{An:2013zwz,Ahn:2012nd,Abe:2012tg}. The most precise
determination of $\sin^2 2 \theta_{13}$ was reported by the Daya
Bay collaboration \cite{An:2013zwz}:
\be
\sin^2 2 \theta_{13} = 0.090^{+0.008}_{-0.009} \;.
\ee
Given the uncertainty in the T2K result, Eq.~(\ref{eq:T2Kth13}), the 
difference between the values of $\sin^2 2 \theta_{13}$ obtained
in the T2K and Daya Bay experiments does not seem to be irreconcilable
and the most natural explanation of this difference can be attributed 
to setting $\delta = 0$ and $\theta_{23} = \pi / 4$.
Indeed, the global analyses of the neutrino oscillation data,
including the data from T2K and Daya Bay,
performed in \cite{Capozzi:2013csa,GonzalezGarcia:2012sz} found 
a hint for non-zero value of $\delta$ and for a deviation of
$\theta_{23}$ from $\pi / 4$:
for the best fit values of $\delta$ and 
$\sin^2 \theta_{23}$ the authors of \cite{Capozzi:2013csa} obtained 
$\delta \simeq 3 \pi / 2$ and $\sin^2 \theta_{23} = 0.42-0.43$.
Similar results were obtained in \cite{GonzalezGarcia:2012sz}.
\\
On-going and future neutrino experiments \cite{Abe:2013hdq,Ayres:2004js,Bass:2013vcg} have the physics potential to improve the 
data on the leptonic CP violation phase $\delta$ and thus to test the indications
for $\delta \sim 3 \pi / 2$ found in the global analyses \cite{Capozzi:2013csa,GonzalezGarcia:2012sz}.

In this letter we would like to entertain a different possibility,
namely, that the difference between the values of $\sin^2 2 \theta_{13}$
found in the T2K experiment for $\delta = 0$, $\theta_{23} = \pi / 4$, etc., 
and in the Daya Bay experiment 
is due to the presence of new physics in the neutrino sector.
More specifically, we consider the effects of 
non-standard neutrino interactions (NSI) \cite{LWNSI78,Grossman:1995wx}
on the $\bar \nu_e \to \bar \nu_e$
and $\nu_\mu \to \nu_e$ oscillation probabilities 
and show how the values 
obtained in the two experiments can be reconciled.

\section{Basic formalism}

\label{Sec:NSI}
In what follows we consider the analytic treatment of 
Non Standard Interactions (NSI) as described in \cite{Kopp:2007ne},
where it was assumed that NSI can affect both neutrino 
production and detection processes. Matter effects 
can be safely neglected in the 
 $\bar \nu_e \to \bar \nu_e$ and $\nu_\mu \to \nu_e$ oscillation probabilities,
 relevant for the interpretation of the Daya Bay and T2K data of interest.

Effects of NSI can appear at low energy through 
unknown couplings $\varepsilon_{\alpha\beta}$, 
generated after integrating out heavy
degrees of freedom. These new couplings can affect neutrino 
production $s$ and detection $d$ \cite{Grossman:1995wx}, so
the neutrino states are a superposition of 
the orthonormal flavor eigenstates $|\nu_e \rangle$,  
$|\nu_\mu \rangle$ and $|\nu_\tau \rangle$
\cite{Ohlsson:2013nna,Ohlsson:2012kf,Meloni:2009cg}:
\begin{eqnarray}\label{eq:s}
|\nu^{\rm s}_\alpha \rangle & = &  |\nu_\alpha \rangle +
\sum_{\beta=e,\mu,\tau} \varepsilon^{ s}_{\alpha\beta}
|\nu_\beta\rangle   = \big[ (1 + \varepsilon^{ s}) |\nu \rangle \big]_{\alpha} \,, \\
\langle \nu^{\rm d}_\beta| & = &\langle
 \nu_\beta | + \sum_{\alpha=e,\mu,\tau}
\varepsilon^{ d}_{\alpha \beta} \langle  \nu_\alpha  | =  \big[ \langle \nu
|  (1 + \varepsilon^{ d}) \big]_{\beta}
\,. \label{eq:d}
\end{eqnarray}
%%%%%%%%%%%%%%%%%%%%%%%%%%%%%%%
The oscillation probability can be obtained by squaring the amplitude $\bra{\nu^d_\beta} e^{-i H L} \ket{\nu^s_\alpha}$:
\begin{align}
  P_{\nu^s_\alpha \rightarrow \nu^d_\beta}
    &= |\bra{\nu^d_\beta} e^{-i H L} \ket{\nu^s_\alpha}|^2 \nonumber\\
    &= \big| (1 + \eps^d)_{\gamma\beta} \, \big( e^{-i H L} \big)_{\gamma\delta}
             (1 + \eps^s)_{\alpha\delta} \big|^2 .          \nonumber
  \label{eq:P-ansatz}
\end{align}
%%%%%%%%%%%%%%%%%%%%%%%%%%%%%%%

Since the parameters $\eps_{e\alpha}^s$ and  $\eps_{\alpha e}^{d}$ receive contributions from 
the same higher dimensional operators, 
one can constrain them by the relation:
\be
\eps_{e\alpha}^s = \eps_{\alpha e}^{d*} 
\equiv  \eps_{e\alpha} e^{{\rm i}  \, \phi_{e\alpha}}\;,
\label{eq:asseps}
\ee
$\eps_{e\alpha}$ and $\phi_{e\alpha}$ being the modulus and 
 the argument of $\eps_{e\alpha}^s$.
%%%%%%%%%%%%%%%%%%%%%%%%%%%%%%%
For $\eps_{\alpha\beta}$ 
there exist model independent bounds derived in \cite{enrique}, which at 90\% C.L. read:
\bea
\label{limitienr}
& \varepsilon_{ee} < 0.041, \quad \varepsilon_{e\mu} < 0.025, 
\quad \varepsilon_{e\tau} < 0.041 \;, \nonumber \\
& |\varepsilon_{\mu e}^{s,d}| < 0.026, \quad |\varepsilon_{\mu \mu}^{s,d}| < 0.078, 
\quad |\varepsilon_{\mu \tau}^{s,d}| < 0.013 \;, 
\eea
%%%%%%%%%%%%%%%%%%%%%%%%%%%%%%%
%
whereas for the CP violation phases  $\phi_{e\alpha}$  
no constraints have been obtained so far. 
These bounds can be further improved, e.g., by future reactor neutrino experiments \cite{Ohlsson:2013nna} and 
at neutrino factories \cite{Coloma:2011rq}, especially
the bounds on non-diagonal couplings
which are expected to be constrained at the level of ${\cal O}(10^{-3})$.
Recently it was shown in Ref. \cite{Girardi:2014gna} that the bound on $\varepsilon_{ee}$ can be improved by almost an order 
of magnitude by the most recent data of the Daya Bay experiment \cite{An:2013zwz}, i.e. $\varepsilon_{ee} \lesssim 3.6 \times 10^{-3}$
at 90\% confidence level.

In the case of the Daya Bay setup, the relevant features of the
$\bar \nu_e \to \bar \nu_e$ survival probability at the far and near detectors can be already caught 
keeping terms up to $O(\eps)$ in the expansion in the small
couplings $|\eps_{\alpha \beta}^{s,d}|$ and neglecting terms of ${\cal O}(\sdm /\ldm)$ and of
 ${\cal O}(\eps \sin^2 \theta_{13},\sin^3 \theta_{13})$.
 \\
 On the other hand, for the T2K setup,
 the correct dependence on the Dirac phase $\delta$ is reproduced keeping
 the first order terms in $\sdm$, as discussed in \cite{Kopp:2007ne}.
 \\
In the limiting case $\eps_{ee} = 0$ (which is a 
good approximation since 
$|\eps_{ee} \cos \phi_{ee}| < O(10^{-3})$  \cite{Girardi:2014gna}), 
the $\ov \nu_e \rightarrow \ov \nu_e$ survival probability 
can be written for $\delta = 0$ as:
%%%%%%%%%%%%%%%%%%%%%%%%%%
\be
\begin{split}
P(\overline \nu_e \rightarrow \overline \nu_e) & = 1 
 - \sin^2 2 \hat \theta_{13} \sin^2  \left[ \frac{\ldm \, L}{4 E_{\nu}} \right ] \;,
\end{split}
\label{Eq:sin2th13effective}
\ee
%%%%%%%%%%%%%%%%%%%%%%%%%%
where \cite{Ohlsson:2008gx}
\be
\begin{split}
\sin^2 2 \hat \theta_{13} & = \sin^2 2 \theta_{13}  + 4 \eps_{e\mu}  \sin 2\theta_{13} \sin \theta_{23} \cos 2 \theta_{13}   \cos(\phi_{e\mu}) \\ 
& + 4 \eps_{e\tau} \sin 2\theta_{13} \cos \theta_{23}  \cos 2 \theta_{13}  \cos(\phi_{e\tau} ) \;.\\
\end{split}
\label{Eq:PNSIlimitepsee0}
\ee
%%%%%%%%%%%%%%%%%%%%%%%%%%
\\
\\
The terms involving the 
parameters $\eps_{e\mu}$ and $\eps_{e\tau}$
can affect significantly the determination of 
the reactor angle $\theta_{13}$, as pointed out in \cite{Girardi:2014gna,Ohlsson:2008gx}. 
Depending on the phases $\phi_{e\mu}$ and $\phi_{e\tau}$, 
relatively large values of $\eps_{e\mu}$ and $\eps_{e\tau}$ 
can lead to smaller (for $\phi_{e\mu} = \phi_{e\tau} \simeq 0$), 
equal (for $\phi_{e\mu} \simeq \phi_{e\tau} + \pi$ and $\eps_{e\mu} \simeq \eps_{e \tau}$) 
or larger (for $\phi_{e\mu} = \phi_{e\tau} \simeq \pi$) 
values of $\theta_{13}$ than those obtained 
in the standard case of absence of NSI.

The oscillation probability $P(\nu_{\mu} \rightarrow \nu_e)$
relevant for the interpretation of the T2K data
on $\sin^2 2 \theta_{13}$, 
can be written for $\delta = 0$, 
$\Delta m^2_{21}/|\Delta m^2_{31} \ll 1$ and
taking into account the NSI as:
%%%%%%%%%%%%%%%%%%%%%%%%%%
\be
\begin{split}
P(\nu_{\mu} \rightarrow \nu_e) & \simeq \sin^2\theta_{23} \sin^22\theta_{13} \sin^2 \frac{\Delta m^2_{31}L}{4E}  + P_0 + P_1 \;,\\
\end{split}
\ee
%%%%%%%%%%%%%%%%%%%%%%%%%%
%
where $P_0$ and $P_1$ include respectively the zero and 
the first order contributions of the NSI,
derived for 
$\sdm L /(4 E_{\nu}) \ll 1$.
Indeed, for the neutrino energy 
of $E_{\nu} = 0.1$ GeV we have:
$\sdm L /(4 E_{\nu}) = 2.7 \times 10^{-4}$ for 
$L = 0.28$ km, % $E_{\nu} = 0.1$ GeV 
and $\sdm L /(4 E_{\nu}) = 0.28$ for $L = 295$ km.
% , $E_{\nu} = 0.1$ GeV. 
\\
Using the constraints given in Eq.~(\ref{eq:asseps}) and defining 
$\eps^{s,d}_{\alpha\beta} = |\eps^{s,d}_{\alpha\beta}|
\exp(i \phi^{s,d}_{\alpha\beta})$, we get:
\be
\begin{split}
P_0 = &-4 |\eps^s_{\mu e}| \sin \theta_{13} \sin \theta_{23} \cos (\phi^s_{\mu e} )  \sin^2 \left[ \dfrac{\Delta m^2_{31} \, L}{4 E_{\nu}} \right ] \\
& -4 |\eps^s_{\mu e}| \sin \theta_{13} \sin \theta_{23} \sin (\phi^s_{\mu e} )  \sin \left[ \dfrac{\Delta m^2_{31} \, L}{4 E_{\nu}} \right ]   \cos \left[ \dfrac{\Delta m^2_{31} \, L}{4 E_{\nu}} \right ] \\
& -4 \eps_{e \mu} \sin \theta_{13} \sin \theta_{23}  \cos (\phi_{e \mu} ) \cos 2 \theta_{23} \sin^2 \left[ \dfrac{\Delta m^2_{31} \, L}{4 E_{\nu}} \right ]  \\
& -4 \eps_{e \mu} \sin \theta_{13}  \sin \theta_{23}  \sin (\phi_{e \mu} ) \sin \left[ \dfrac{\Delta m^2_{31} \, L}{4 E_{\nu}} \right ]   \cos \left[ \dfrac{\Delta m^2_{31} \, L}{4 E_{\nu}} \right ]  \\
& + 8 \eps_{e \tau} \sin \theta_{13} \sin^2 \theta_{23} \cos \theta_{23} \cos (\phi_{e \tau} ) \sin^2 \left[ \dfrac{\Delta m^2_{31} \, L}{4 E_{\nu}} \right ] + O(\eps \sin^2 \theta_{13})  + O(\eps^2) \;,\\
\end{split}
\ee

\be
\begin{split}
P_1 = & -|\eps^s_{\mu e}| \sin 2 \theta_{12} \cos \theta_{23} \sin \phi^s_{\mu e} \frac{\sdm L}{2 E_{\nu}}  \\
& + 2 \eps_{e \mu} \sin 2 \theta_{12} \sin^2 \theta_{23} \cos\theta_{23}  \cos \phi_{e \mu} \frac{\sdm L}{4 E_{\nu}} \sin \left[ \dfrac{\Delta m^2_{31} \, L}{2 E_{\nu}} \right ] \\
& +  \eps_{e \mu} \sin 2 \theta_{12} \cos \theta_{23} \sin \phi_{e \mu} \frac{\sdm L}{2 E_{\nu}} \left[1 - 2 \sin^2 \theta_{23} \sin^2 \left[ \dfrac{\Delta m^2_{31} \, L}{4 E_{\nu}} \right ] \right]\\
& + 2 \eps_{e \tau} \sin 2 \theta_{12} \sin \theta_{23} \cos^2 \theta_{23} \cos \phi_{e \tau}   \frac{\sdm L}{4 E_{\nu}} \sin\left[ \dfrac{\Delta m^2_{31} \, L}{2 E_{\nu}} \right ]\\
& -  2 \eps_{e \tau} \sin 2 \theta_{12} \sin \theta_{23} \cos^2 \theta_{23} \sin \phi_{e \tau} \frac{\sdm L}{2 E_{\nu}} \sin^2\left[ \dfrac{\Delta m^2_{31} \, L}{4 E_{\nu}} \right ]  \\
& + O\left(\eps \sin \theta_{13} \frac{\sdm L}{4 E_{\nu}} \right)  + O(\eps^2) \,.\\
\end{split}
\ee
In the previous equations, the $P_0$ term encodes the correlations between $\theta_{13}$ and the new physics parameters so,
as in the Daya Bay case, we expect a significant impact of degeneracies on the determination of the reactor angle.
The term $P_1$ is subleading, whose magnitude is controlled by $\sdm L /(4 E_{\nu}) \ll 1$.

\section{Fit results}

As we can see from the previous formulae, the parameter space for NSI relevant for our analysis 
consists of six parameters, the moduli $\eps_{e\mu}$,
$\eps_{e\tau}$, $\eps^s_{\mu e}$ and the phases
$\phi_{e\mu}$, $\phi_{e\tau}$, $\phi^s_{\mu e}$.
However, for the illustrative purposes of the present study 
it is sufficient to consider  
a smaller parameter space with just two independent NSI parameters, specified below.
We consider two different scenarios: one in which a large 
$\sin^2 2\theta_{13} = 0.14$ ($\sin^2 2\theta_{13} = 0.17$) for NO (IO) 
can be reconciled with both the Daya Bay and 
T2K data and a second where we assume that $\sin^2 2\theta_{13} = 0.09$.

\subsection{\texorpdfstring{The case of  $\sin^2 2 \theta_{13} = 0.14 \, (0.17)$}{The case of large theta13}}

In this case we reduced the parameter space assuming:
\be
\eps = \eps_{e\mu} = \eps_{e\tau} = \eps^s_{\mu e} , \quad 
\phi = \phi_{e\mu} = \phi_{e\tau} , \quad \phi^s_{\mu e} = 0.
\label{eq:largeth13ass}
\ee
The choice of the parameter space is not completely arbitrary. 
For the large $\theta_{13}$ case we need relatively large NSI effects to obtain 
an effective reactor angle satisfying the Daya Bay measurement.

\begin{figure}[h!]
  \begin{center}
   \subfigure
 {\includegraphics[width=6cm]{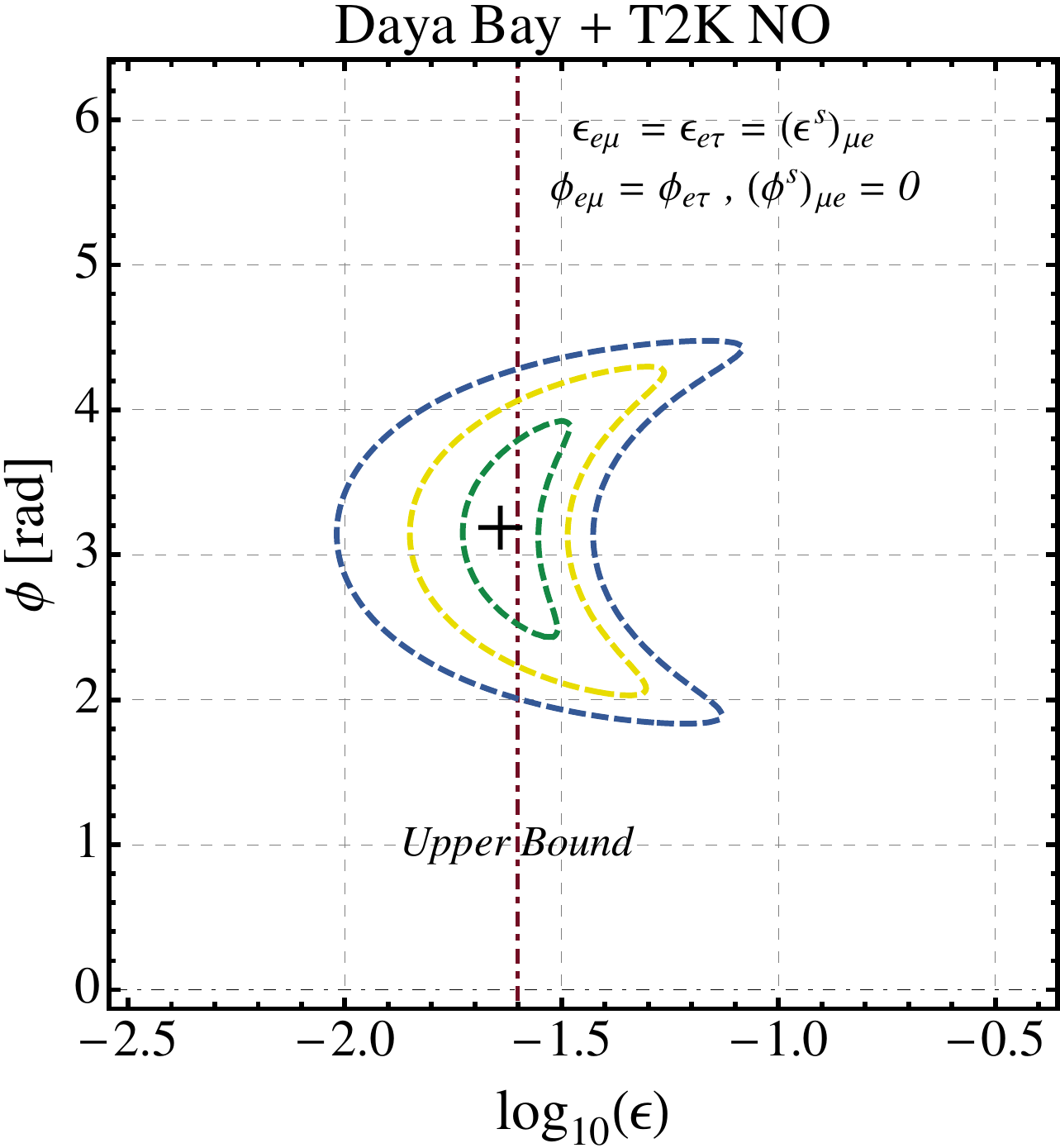}}
    {\includegraphics[width=6.14cm]{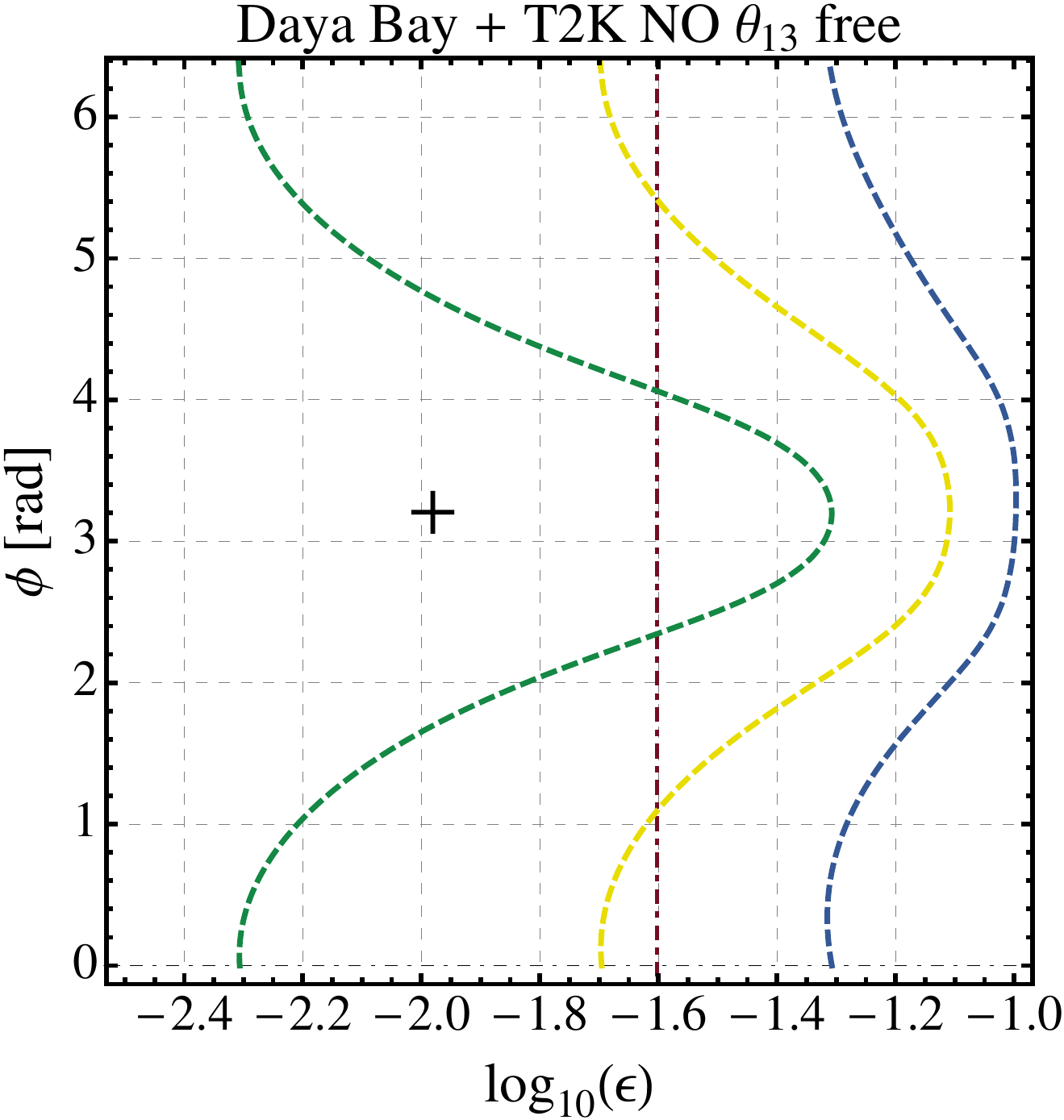}}
 \vspace{5mm}
      \end{center}
\vspace{-1.0cm} \caption{ \it \label{fig:epsphiNO} 
Allowed regions in the $\phi - \log_{10} ( \eps )$ plane, 
where $\eps$ and $\phi$ are respectively the modulus and the phase of the NSI parameter,
at 1$\sigma$, 2$\sigma$ and 3$\sigma$ confidence level (C.L.) for 1 dof fitting the data of the Daya Bay
and the T2K experiments in the case of NSI with NO. The best fit points correspond to the crossed points.
The vertical lines are at $\log_{10} \eps = \log_{10} 0.025$.
}
\end{figure}

\begin{figure}[h!]
  \begin{center}
    \subfigure
  {\includegraphics[width=6cm]{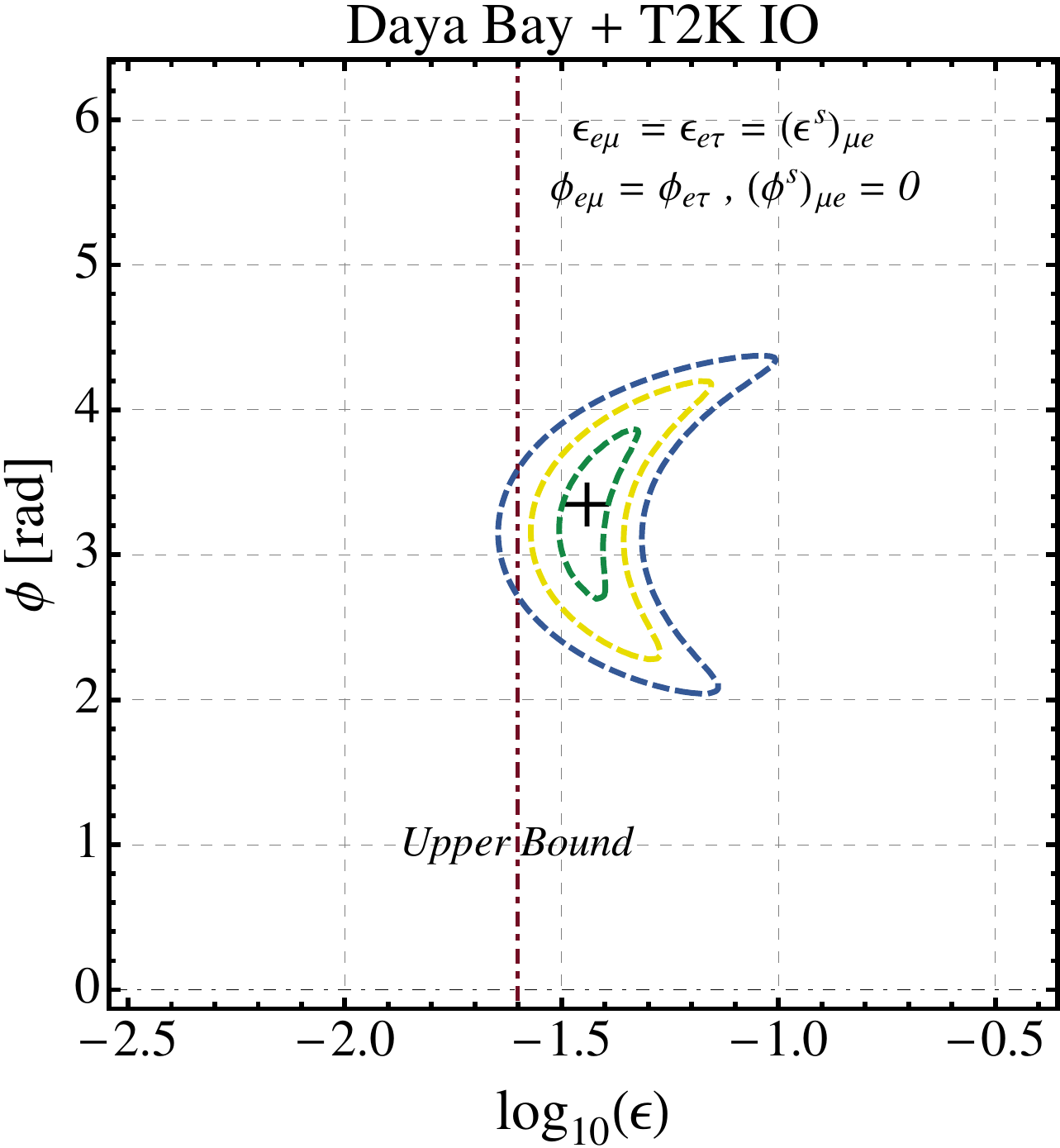}}
 \vspace{5mm}
     {\includegraphics[width=6.14cm]{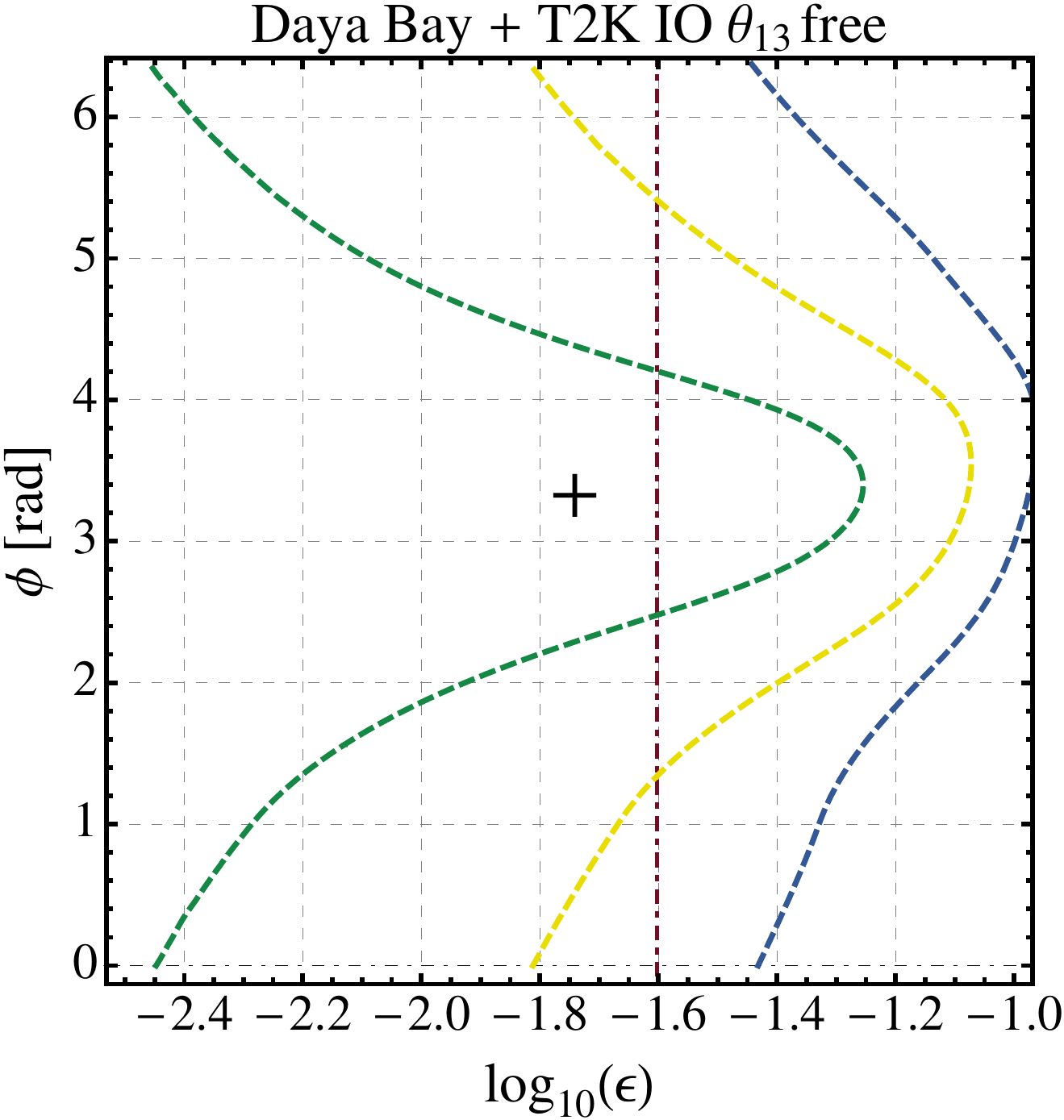}}
 \vspace{5mm}
      \end{center}
\vspace{-1.0cm} \caption{ \it \label{fig:epsphiIO} 
Same as Fig.~\ref{fig:epsphiNO} but for IO. 
}
\end{figure}

\begin{figure}[h!]
   \begin{center}
    \subfigure
 {\includegraphics[width=6cm]{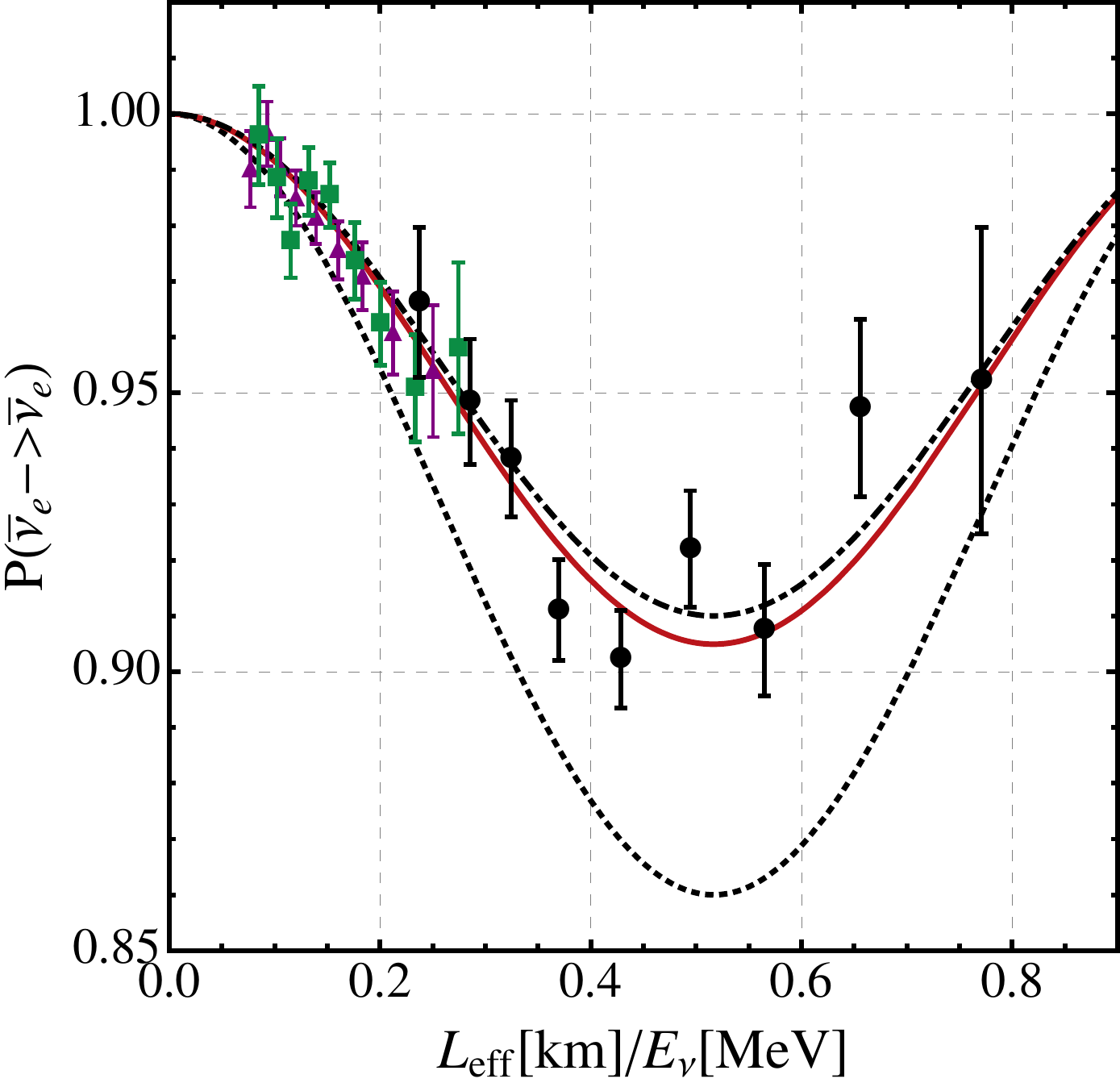}}
  \vspace{5mm}
   \subfigure
   {\includegraphics[width=5.67cm]{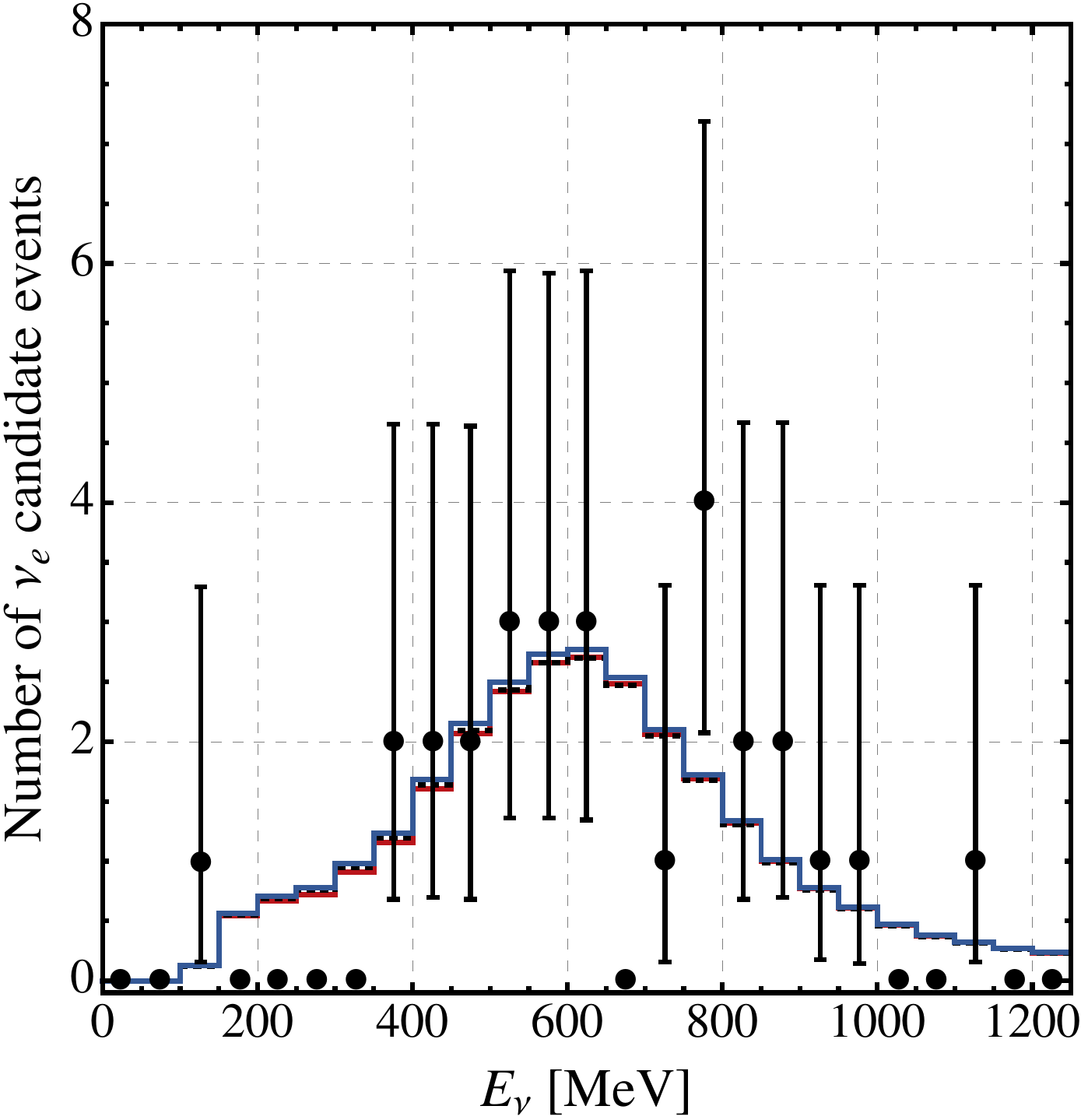}}
  \vspace{5mm}
       \end{center}
 \vspace{-1.0cm} \caption{ \it \label{fig:SpectrumNO} Left Panel. Oscillation
 probability $P(\ov \nu_e \rightarrow \ov \nu_e)$ as a function of $L_{\rm eff}/E_{\nu}$ for the NSI model (solid red line)
 and the SR with $\sin^2 2\theta_{13} = 0.140$ (dashed black line)
 and with $\sin^2 2\theta_{13} = 0.090$ (dot-dashed black line). 
 The triangular,  square and  circular data points refer to the 
 EH1, EH2 and EH3 locations and have been taken from \cite{An:2013zwz}.
 Right Panel. Number of $\nu_{e}$ candidate events as a function
 of the neutrino energy for the NSI model (solid red line),
 the SR (dashed black line) and the T2K best fit curve (solid blue line),
 the three curves being almost superimposed.
 The T2K data and the errors have been taken
 from \cite{Abe:2013hdq}. See the text for further details.
 }
 \end{figure}

In Fig.~\ref{fig:epsphiNO} we show the best fit points and the
1, 2 and 3$\sigma$ confidence level regions for 1 degree of freedom (dof)
after performing a combined fit to the Daya Bay \cite{An:2013zwz} and T2K \cite{Abe:2013hdq} data
(see the Appendix \ref{App:A} for a detailed description of the fitting procedure).
In the left panel of Fig.~\ref{fig:epsphiNO}
we fixed $\sin^2 \theta_{12} = 0.306$, 
$\sdm = 7.6 \times 10^{-5} \; {\rm eV}^2$, $\sin^2 \theta_{23} = 0.5$, 
$|\Delta m^2_{32} | = 2.4 \times 10^{-3} \; {\rm eV}^2$, $\delta = 0$
and  $\sin^2 2\theta_{13} = 0.140$, whereas in the right panel  we allowed $\theta_{13}$
to vary freely, using the mean value and the 1$\sigma$ error as determined in the T2K
experiment, $\sin^2 2 \theta^{T2K}_{13} = 0.140 \pm 0.038$.
\\
Results in the case of inverted hierarchy are shown in Fig.~\ref{fig:epsphiIO}; 
the procedure adopted is the same as the one
used to obtain Fig.~\ref{fig:epsphiNO}, the only difference being that
the fixed value of the reactor angle is now at $\sin^2 2\theta_{13} = 0.170$
and that, when $\theta_{13}$ is left free to vary, 
we used $\sin^2 2 \theta^{T2K}_{13} = 0.170 \pm 0.045$. 
\\
As it can be seen, in the left panels of Figs.~\ref{fig:epsphiNO} and \ref{fig:epsphiIO},
the same value of $\theta_{13}$ can give a good description of both 
Daya Bay and T2K data under the hypothesis of relatively large 
$\varepsilon$ and for a phase $\phi$ which is almost CP conserving.
\\
Since we are adopting the preferred T2K value of $\theta_{13}$, it is necessary to allow for relatively
large NSI couplings to reconcile $\sin^2 2 \theta_{13} = 0.14$ ($\sin^2 2 \theta_{13} = 0.17$)
with the Daya Bay event distribution.
On the other hand, our choice of couplings, Eq.~(\ref{eq:largeth13ass}), does not 
lead to a significant change of the fit to the 
T2K data.
\\
In the case we vary freely $\theta_{13}$ (see Appendix \ref{App:A} for details)
the sensitivity to $\eps$ is significantly reduced (with the smallest statistical sensitivity
at $\phi\sim \pi$), due to the strong correlation between $\theta_{13}$
and the NSI parameters \cite{Girardi:2014gna}.
That means that there exist a vast parameter space for  NSI for 
which the data can be fitted simultaneously at the price of changing accordingly the value of $\theta_{13}$. To give an example, 
at the best fit point we get: $\sin^2 2 \theta_{13} = 0.113$ ($\sin^2 2 \theta_{13} = 0.130$) for the NO (IO) spectrum.
%%%%%%%%%%%%%%%%%%%%%%%%%%%%%%%%%%%%%%%%%%%%%%%%%%%%%%%%%
%%%%%%%%%%%%%%%%%%%%%%%%%%%%%%%%%%%%%%%%%%%%%%%%%%%%%%%
%
%
%
%%%%%%%%%%%%%%%%%%%%%%%%%%%%%%%%%%%%%%%%%%%%%%%%%%%%%%%%%
%%%%%%%%%%%%%%%%%%%%%%%%%%%%%%%%%%%%%%%%%%%%%%%%%%%%%%%
%
%
\\
The values of $\eps\,,\, \phi$ at the NO(IO) best fit
point are given in Table \ref{tab:bfLth13}. 
\begin{table}[h]
\begin{center}
\begin{tabular}{|c | c  | c |}
\hline
\bf $ (\log_{10} \eps \,,\, \phi)$ best fit & \bf left panel & \bf right panel  \\
\hline
\bf NO &   $(-1.64\,,\,3.18)$ &  $(-1.98\,,\,3.20)$ \\
\bf IO &   $(-1.44\,,\, 3.34)$ &  $(-1.74\,,\,3.32)$ \\
\hline
\end{tabular}
\caption{\it Best fit points for the $(\log_{10} \eps \,,\, \phi)$ parameters obtained in our analysis.
NO refers to Fig.~\ref{fig:epsphiNO}, IO to Fig.~\ref{fig:epsphiIO}.}
\label{tab:bfLth13}
\end{center}
\end{table}
We notice that 
the confidence level regions are slightly shifted to the left (right)
if instead of the assumption in Eq.~(\ref{eq:largeth13ass})
we impose: $\eps_{e\mu} = 2 \, \eps_{e\tau} = \eps^s_{\mu e}$, 
$\phi_{e\mu} = \phi_{e\tau}$ and $\phi^s_{\mu e} = 0$  
($\eps_{e\mu} = \eps_{e\tau} = \eps^s_{\mu e}$, 
$\phi_{e\mu} = \phi_{e\tau}$ and $\phi^s_{\mu e} = \pi/2$).\\
To demonstrate that for the obtained values of the NSI parameters 
one can
describe both the Daya Bay and T2K 
results, including the spectra, 
in the Left Panel of Fig.~\ref{fig:SpectrumNO} we show the oscillation probability $P(\ov \nu_e \rightarrow \ov \nu_e)$ as
 a function of $L_{\rm eff}/E_{\nu}$ \cite{An:2013zwz} for the NSI model (solid red line) for NO spectrum
 and in the absence of NSI ("standard result" (SR)) (dotted black line); the mixing parameters are fixed as follows:
 $\sin^2 \theta_{12} = 0.306$, 
 $\sdm = 7.6 \times 10^{-5} \; {\rm eV}^2$, $\sin^2 \theta_{23} = 0.5$, 
 $|\Delta m^2_{32} | = 2.4 \times 10^{-3} \; {\rm eV}^2$, $\delta = 0$
 and  $\sin^2 2\theta_{13} = 0.140$,
 $\eps_{e\mu} = \eps_{e\tau} = \eps^s_{\mu e} = 10^{-1.64}$, $\phi_{e\mu} = \phi_{e\tau} = 3.18$
 and $\phi^s_{\mu e} = 0$.
 The triangular,  square and  circular data points refer to the 
 EH1, EH2 (near detectors) and EH3 (far detector) Daya Bay locations and have been taken from \cite{An:2013zwz}.
 The Right Panel of Fig.~\ref{fig:SpectrumNO} has been obtained using the same
 values for the standard oscillation and NSI parameters and shows the
 number of candidate events in the appearance channel of the T2K experiment.
 The SR result with $\sin^2 2\theta_{13} = 0.090$ is shown with the dot-dashed line in the
 left panel and the T2K best fit curve is represented with the blue line in the right panel.
 As it is clear from these figures, the Daya Bay and the T2K spectral data are well reproduced.
 
% %%%%%%%%%%%%%%%%%%%%%%%%%%%%%%%%%%%%%%%%%%%%%%%%%%%%%%%%%
%%%%%%%%%%%%%%%%%%%%%%%%%%%%%%%%%%%%%%%%%%%%%%%%%%%%%%%

\subsection{\texorpdfstring{The case of $\sin^2 2 \theta_{13} = 0.09$}{The case of small theta13}}

\begin{figure}[h!]
  \begin{center}
  \subfigure
    {\includegraphics[width=6cm]{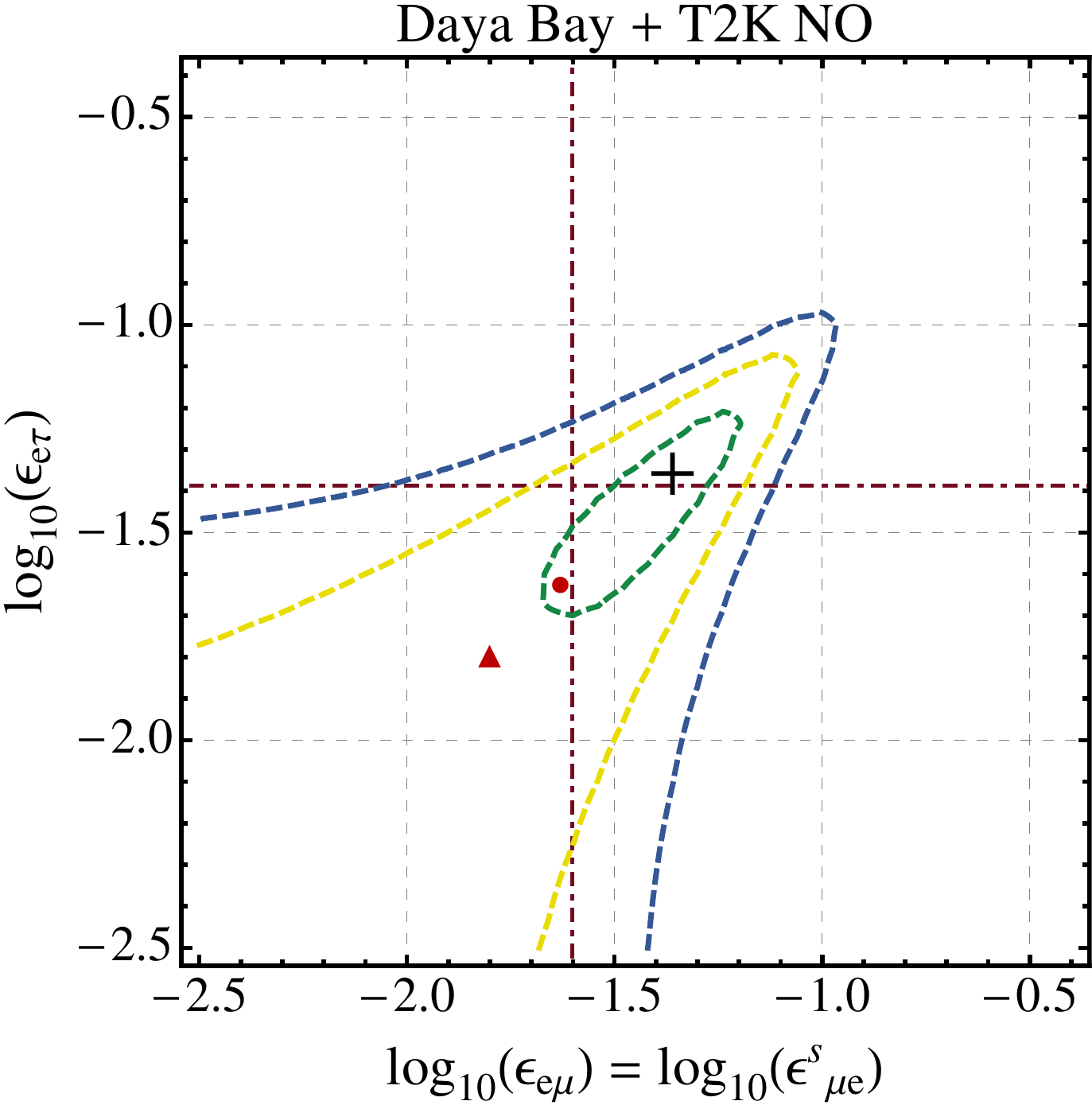}}
 \vspace{5mm}
  \subfigure
      {\includegraphics[width=6cm]{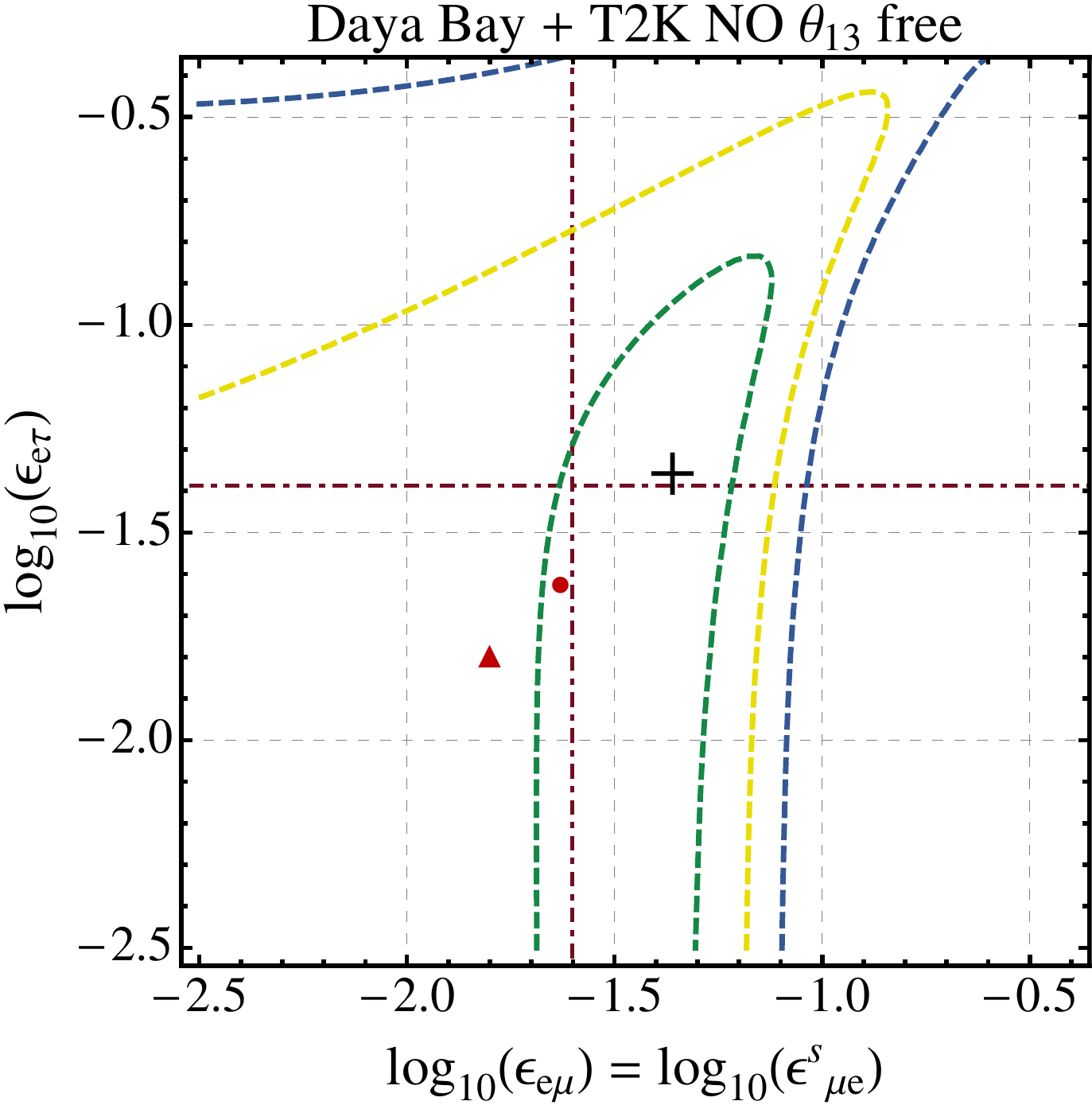}}
 \vspace{5mm}
      \end{center}
\vspace{-1.0cm} \caption{\it \label{fig:epsphiSNOeps} 
Allowed regions in the $\log_{10} \eps_{e\tau} - \log_{10} \eps$ plane, 
where $\eps$ and $\phi$ are respectively the modulus and the phase of the NSI parameter,
at 1$\sigma$, 2$\sigma$ and 3$\sigma$ confidence level (C.L.) for 1 dof fitting the data of the Daya Bay
and the T2K experiments in the case of NSI with NO. 
The best fit points correspond to the crossed points; the vertical lines are at $\log_{10} \eps = \log_{10} 0.025$,
the horizontal lines at $\log_{10} \eps_{e\tau} = \log_{10} 0.041$.
The circular and triangular points are at $(\log_{10} \eps \,,\, \log_{10} \eps_{e\tau}) = (-1.63\,,\,-1.63) \,,\,(-1.80\,,\,-1.80)$,
respectively.  
}
\end{figure}

\begin{figure}[h!]
   \begin{center}
    \subfigure
 {\includegraphics[width=6cm]{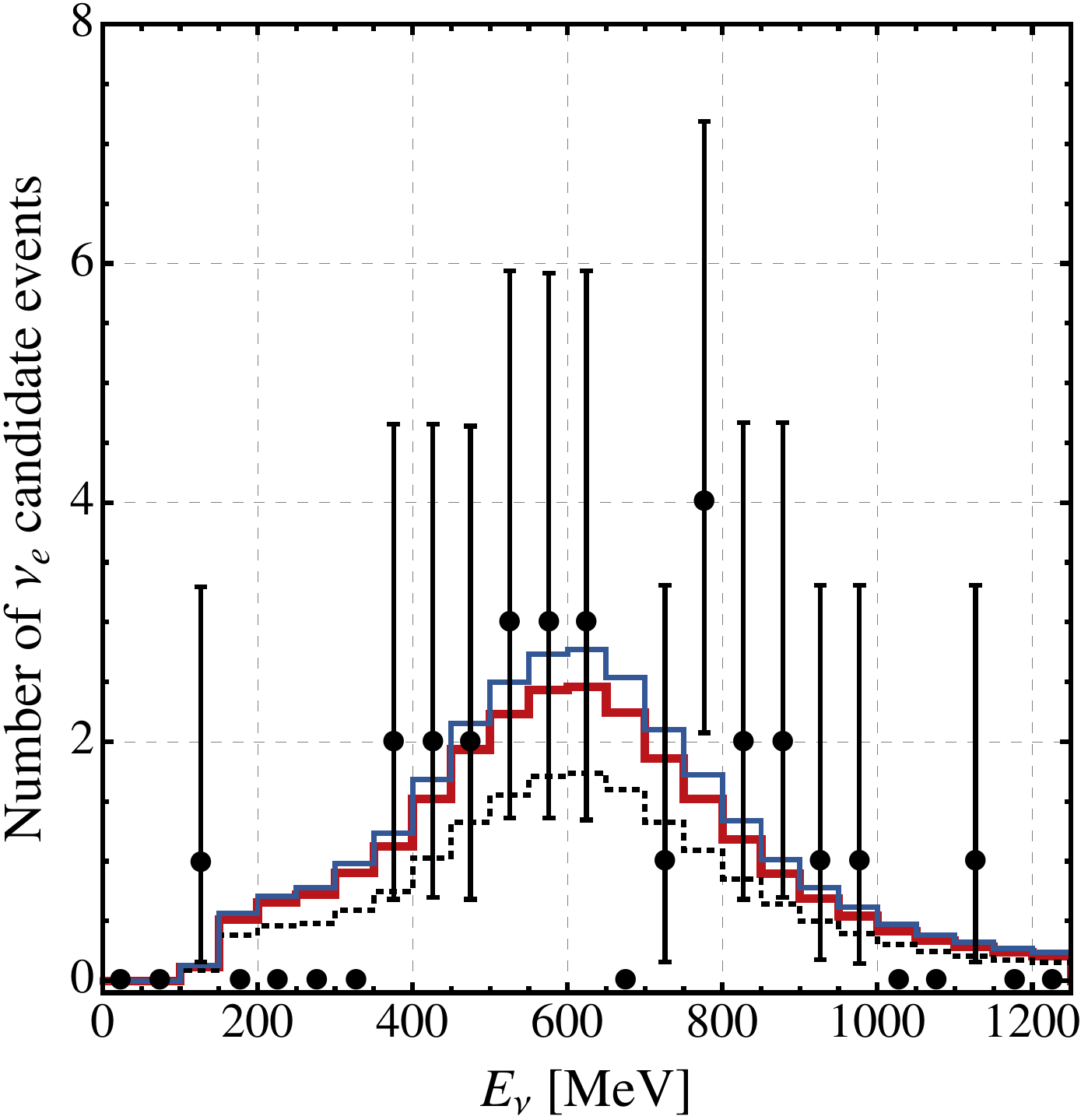}}
  \vspace{5mm}
   \subfigure
   {\includegraphics[width=6cm]{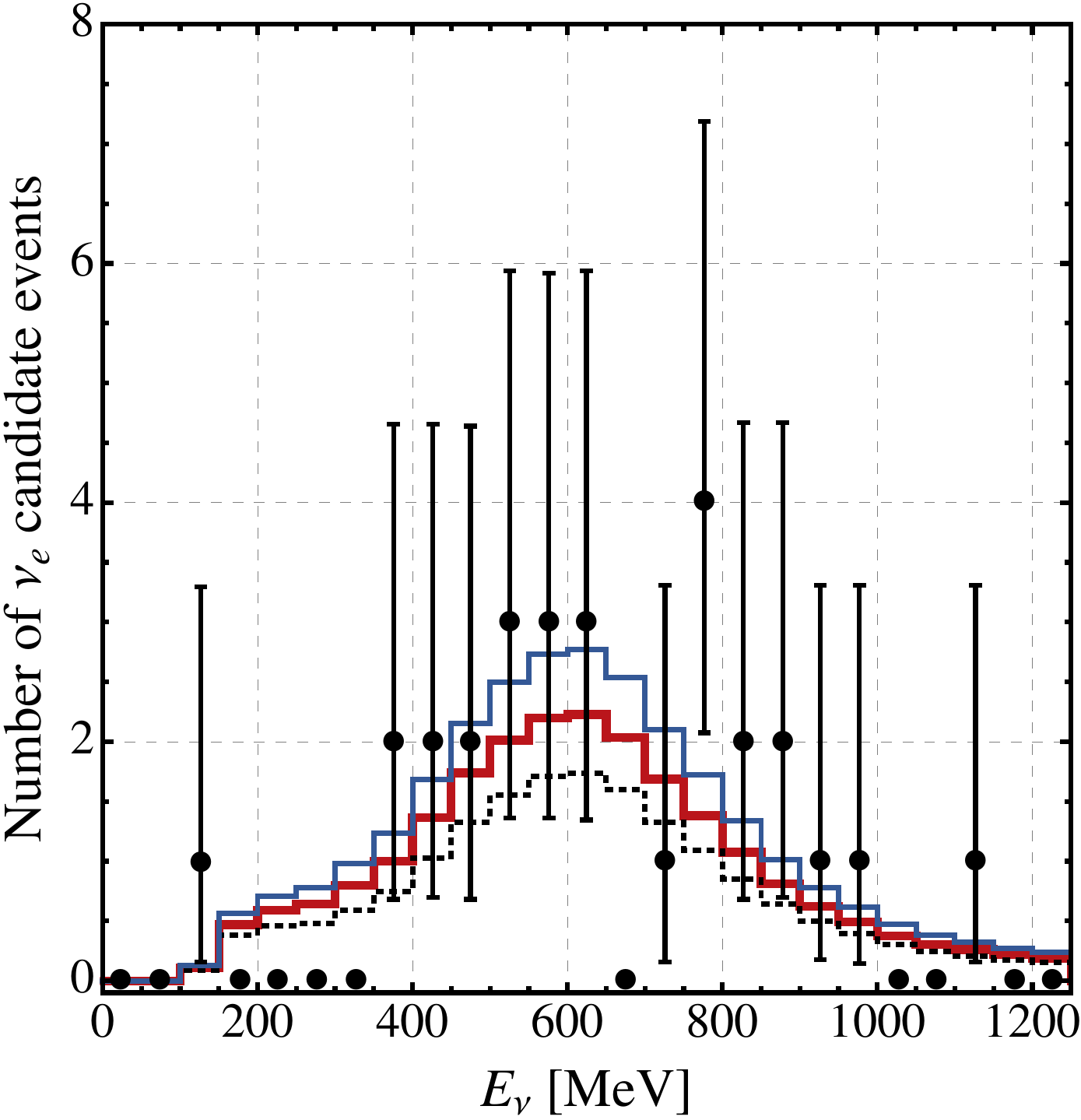}}
  \vspace{5mm}
       \end{center}
 \vspace{-1.0cm} \caption{\it \label{fig:SpectrumSNOeps} 
 Left Panel. Number of $\nu_{e}$ candidate events as a function
 of the energy for the NSI model (solid red line) with $(\log_{10} \eps \,,\, \log_{10} \eps_{e\tau}) = (-1.63\,,\,-1.63)$,
 the SR (dashed black line) and the T2K best fit curve (solid thin blue line). 
 The T2K data and the errors have been taken
 from \cite{Abe:2013hdq}. 
 Right Panel. As in the Left Panel but using
 $(\log_{10} \eps \,,\, \log_{10} \eps_{e\tau}) = (-1.80\,,\,-1.80)$.
 See the text for further details.
 }
 \end{figure}

In the case of small $\theta_{13}$ we reduced the parameter space assuming:
\be
\eps = \eps_{e\mu} = \eps^s_{\mu e}, \quad \eps_{e\tau} \neq 0 , \quad 
\phi_{e\mu} = \phi^s_{\mu e} = \pi, \quad \phi_{e\tau} = 0.
\label{eq:largeth13ass2}
\ee
In the case of small $\theta_{13}$
the choice in Eq. (13) is dictated by the need of 
minimizing the NSI effects in the $\bar\nu_e \rightarrow \bar\nu_e$ survival probability, so that the results of the Daya Bay fit remain unaffected.
In the Left Panel of Fig.~\ref{fig:epsphiSNOeps} we show the best fit points and the
1, 2 and 3$\sigma$ confidence level regions for 1 dof
after performing a combined fit to the Daya Bay and to the
T2K data for NO fixing $\sin^2 \theta_{12} = 0.306$, 
$\sdm = 7.6 \times 10^{-5} \; {\rm eV}^2$, $\sin^2 \theta_{23} = 0.5$, 
$|\Delta m^2_{32} | = 2.4 \times 10^{-3} \; {\rm eV}^2$, $\delta = 0$
and  $\sin^2 2\theta_{13} = 0.09$.
In the Right Panel of Fig.~\ref{fig:epsphiSNOeps} we allowed
$\theta_{13}$ to vary freely.
\\
We do not show the results for the IO spectrum,
because, under the assumptions made for the parameter space, Eq.~\ref{eq:largeth13ass2},
they are the same as in the NO case.
\\
In contrast to the large $\theta_{13}$ case, 
in order to reconcile the Daya Bay and the T2K 
spectral data 
requires that the phase $\phi_{e\mu}$ and $\phi_{e\tau}$ are
related through $\phi_{e\mu} \simeq \phi_{e\tau} - \pi$. This ensures
that sizeable NSI effects do not spoil the Daya Bay  measurement
of the reactor angle when $\eps_{e\mu} \sim \eps_{e\tau}$: 
in fact, $P(\ov \nu_e \rightarrow \ov \nu_e)$ is reduced essentially to the standard expression and no 
significant effect has to be expected from the NSI parameters
at leading order. 
On the other hand, it is clear that relatively large values of $\varepsilon$ are needed to fit the T2K data. 
%
%
%
%%%%%%%%%%%%%%%%%%%%%%%%%%%%%%%%%%%%%%%%%%%%%%%%%%%%%%%%%
%%%%%%%%%%%%%%%%%%%%%%%%%%%%%%%%%%%%%%%%%%%%%%%%%%%%%%%
%
%
%
\\
We give in Table \ref{tab:bfSth13eps} the best fit points we obtained in our analysis
for Fig.~\ref{fig:epsphiSNOeps}. Notice that they are close to
the current upper limits, reported with dot-dashed lines.
\begin{table}[h]
\begin{center}
\begin{tabular}{|c | c | c  |}
\hline
\bf  best fit & \bf left panel & \bf right panel   \\
\hline
$ (\log_{10} \eps \,,\, \log_{10} \eps_{e\tau})$ & $(-1.36\,,\,-1.36)$ & $(-1.36\,,\,-1.36)$ \\
\hline
\end{tabular}
\caption{\it Best fit points for the $(\log_{10} \eps\,,\, \log_{10} \eps_{e\tau})$ parameters obtained in our analysis.}
\label{tab:bfSth13eps}
\end{center}
\end{table}
\\
Finally, in Fig.~\ref{fig:SpectrumSNOeps} we show 
the number of candidate events in the appearance channel of the T2K experiment
(with mixing parameters fixed at the values discussed below Eq.~\ref{eq:largeth13ass2}).\\
 Since the best fit points are outside the current 90\% C.L. bounds
 on the NSI parameters, we show the spectra for two points within
 the NSI bounds: one point is located in the 1$\sigma$ region, while
 the second is located in the 2$\sigma$ region (see Fig.~\ref{fig:epsphiSNOeps}).
 In the Left Panel of Fig.~\ref{fig:SpectrumSNOeps} we fixed
 $(\log_{10} \eps \,,\, \log_{10} \eps_{e\tau}) = (-1.63\,,\,-1.63)$,
 in the Right Panel  $(\log_{10} \eps \,,\, \log_{10} \eps_{e\tau}) = (-1.80\,,\,-1.80)$. 
 The T2K best fit curve is represented with the blue line.
 As it is clear from these figures, the T2K spectral data are well reproduced.
% 
% 
% 
% 
% %%%%%%%%%%%%%%%%%%%%%%%%%%%%%%%%%%%%%%%%%%%%%%%%%%%%%%%%%
% %%%%%%%%%%%%%%%%%%%%%%%%%%%%%%%%%%%%%%%%%%%%%%%%%%%%%%%
% 

\section{Conclusions}

In the present paper we have analyzed the most recent data
of the Daya Bay \cite{An:2013zwz} and the T2K \cite{Abe:2013hdq} experiments with the aim to
study the possibility that NSI effects
can reconcile the different values of the reactor angle 
reported by the two experiments. We recall that the best fit values of 
$\sin^2 2 \theta_{13}$ found in the experiments, $\sin^2 2 \theta_{13} = 0.090$ \cite{An:2013zwz}
and $\sin^2 2 \theta_{13} = 0.140$ (0.170) \cite{Abe:2013hdq},
differ by a factor 1.6 (1.9) in the case of NO (IO) neutrino mass spectrum.
The T2K result was obtained under the assumptions: i) the Dirac CP violation phase $\delta = 0$, 
ii) the atmospheric neutrino mixing angle $\theta_{23} = \pi/4$,
iii) $\sin^2 \theta_{12} = 0.306$, iv) $\sdm = 7.6 \times 10^{-5} \; {\rm eV}^2$
and v) $|\Delta m^2_{32} | = 2.4 \times 10^{-3} \; {\rm eV}^2$.
Given the uncertainty in the T2K result, the
difference between the values of $\sin^2 2 \theta_{13}$ obtained
in the T2K and Daya Bay experiments does not seem to be irreconcilable and
the most natural explanation can be attributed to setting $\delta = 0$ and
$\theta_{23} = \pi / 4$.
In this Letter we have entertained a different possibility,
namely, that the difference between the values of $\sin^2 2 \theta_{13}$
found in the T2K experiment for $\delta = 0$
and in the Daya Bay experiment are due to the
presence of new physics in the neutrino sector in the form
of non-standard neutrino interactions (NSI).
There are altogether six NSI parameters which can affect the  
$\bar \nu_e \to \bar \nu_e$ and $\nu_\mu \to \nu_e$ oscillation probabilities,
relevant for the interpretation of the Daya Bay and T2K data on $\sin^2 2 \theta_{13}$:
three complex, in general, NSI effective couplings, whose absolute values and phases are
$\eps_{e\mu}$, $\eps_{e\tau}$, $\eps^s_{\mu e}$ and $\phi_{e\mu}$, $\phi_{e\tau}$, $\phi^s_{\mu e}$.
We have considered two extreme cases: one where the true value of $\theta_{13}$ is $\sin^2 2\theta_{13} = 0.140$ 
for NO ($\sin^2 2\theta_{13} = 0.170$ for IO),
and the other where the true value is $\sin^2 2\theta_{13} = 0.090$.
With the aim of finding a minimal model with few new degrees of freedom
for each of the two cases, we have simplified the NSI parameter spaces,
assuming 
$
\eps = \eps_{e\mu} = \eps_{e\tau} = \eps^s_{\mu e} , 
\phi = \phi_{e\mu} = \phi_{e\tau} ,  \phi^s_{\mu e} = 0
$
for the large $\theta_{13}$ case and 
$\eps = \eps_{e\mu} = \eps^s_{\mu e}, \eps_{e\tau} \neq 0 , 
\phi_{e\mu} = \phi^s_{\mu e} = \pi, \phi_{e\tau} = 0$
for the small $\theta_{13}$ one.
All other mixing parameters are fixed to $\sin^2 \theta_{12} = 0.306$, 
$\sdm = 7.6 \times 10^{-5} \; {\rm eV}^2$, $\sin^2 \theta_{23} = 0.5$, 
$|\Delta m^2_{32} | = 2.4 \times 10^{-3} \; {\rm eV}^2$, $\delta = 0$.
We have found that, contrary to the interpretation that $\delta = 0$ is disfavoured in the 
standard case, following from the global analysis of the neutrino oscillation 
data \cite{Capozzi:2013csa,GonzalezGarcia:2012sz},
it is possible 
to find a good agreement with both the hypothesis of large, $\sin^2 2 \theta_{13} = 0.14$ (0.17), and 
small, $\sin^2 2 \theta_{13} = 0.09$,
for $\delta = 0$, in well defined regions of the NSI parameter space.
In a more general situation in which the NSI can affect the neutrino flux in
the near detector and without the restrictions we considered on the parameter space,
it will be possible to reconcile the Daya Bay and T2K data in a bigger
region of the NSI parameter space within the current upper bounds.
\\
Given the relatively low statistics of the T2K $\nu_{\mu} \rightarrow \nu_e$
oscillation data, our results on the possible NSI effects should be considered
as very preliminary.
Future experiments searching the CP violation and/or NSI effects
in neutrino oscillations will certainly provide a critical test of the
possible NSI effects discussed in the present article.

\section*{Acknowledgments}
We acknowledge MIUR (Italy) for
financial support under the program Futuro in Ricerca 2010 (RBFR10O36O). 
 This work was supported in part by the INFN program on
 ``Astroparticle Physics'' and  by the European Union FP7-ITN INVISIBLES
 (Marie Curie Action PITAN-GA-2011-289442-INVISIBLES) (I.G. and S.T.P.).

\appendix

\section{The Daya Bay and the T2K experiments}
\label{App:A}
The Daya Bay experimental setup we take into account \cite{An:2013uza} consists of six antineutrino detectors (ADs)
and six reactors; detailed information on the antineutrino spectra emitted by the nuclear reactors 
and arriving to the detectors can be found in \cite{Mueller:2011nm,Huber:2011wv,talk:DayaBay}. 
For our analysis we used the data set accumulated during 217 days 
reported in \cite{An:2013zwz}, where the detected antineutrino candidates are collected in the far hall, EH3
(far detector), and 
in the near halls EH1, EH2 (near detectors). 

The antineutrino energy $E_{\overline \nu_e}$ is reconstructed
by the prompt energy deposited by the positron $E_{ \rm prompt}$ using 
the approximated relation \cite{An:2013uza}:
$E_{\overline \nu_e} \simeq E_{\rm prompt} + 0.8 \; {\rm MeV}$.
We adopt a Gaussian energy resolution
function of the form:
\be
R^c(E,E^{\prime}) = \frac{1}{\sigma(E) \sqrt{2 \pi}} e^{- \frac{(E - E^{\prime})^2}{2 \sigma^2(E)}} \;.
\label{Eq:res}
\ee 
with $\sigma(E) [\rm MeV] = \alpha \cdot E + \beta \cdot \sqrt{E} + \gamma$ that, for Daya Bay, 
are $(\alpha,\beta,\gamma) = (0,0,0.08)$ MeV.
The antineutrino cross section for the inverse beta decay (IBD) process has been taken
from \cite{Vogel:1999zy}. 
The statistical analysis of the data has been performed using the GLoBES software
\cite{GLOB2} with the $\chi^2$ function defined as \cite{An:2013uza}:
\begin{eqnarray}
 \label{eqn:chispec}
&& \chi^2_{DB}(\theta,\Delta m^2, \vec S,\alpha_r,\varepsilon_d,\eta_d) = \nonumber 
 \sum_{d=1}^{6}\sum_{i=1}^{36}
 \frac{\left[M_i^d -T_i^d \cdot \left(1 +
 \sum_r\omega_r^d\alpha_r + \varepsilon_d  \right) 
  + \eta_d \right]^2}
 {M_i^d + B_i^d } \nonumber \\
&& + \sum_r\frac{\alpha_r^2}{\sigma_r^2}
 + \sum_{d=1}^{6}\left[
\frac{\varepsilon^2_d}{\sigma^2_d}
 + \frac{\eta_d^2}{\sigma_{B_d}^2}
 \right]   + \mbox{Priors} \,,
\end{eqnarray}
%%%%%%%%%%%%%%%%%%%%%%%%%%%%%%%
where $\vec S$ is a vector containing the new physics parameters,
$M^d_i$ are the measured IBD events of the d-th detector ADs
in the i-th bin, $B^d_i$ the corresponding background and $T^d_i = T_i(\theta,\Delta m^2, \vec S)$ are the
theoretical prediction for the rates. The parameter $\omega_r^d$ is the fraction
of IBD contribution of the r-th reactor to the d-th detector AD, determined by
the approximated relation $\omega_r^d \sim L_{rd}^{-2} / (\sum_{r = 1}^6 1/L_{rd}^2 )$,
where $L_{rd}$ is the distance between the d-th detector and the r-th reactor. The parameter $\sigma_d$
is the uncorrelated detection uncertainty ($\sigma_d = 0.2$\%) and $\sigma_{B_d}$ is the background uncertainty of the d-th detector 
obtained using the information given in \cite{An:2013zwz}:
$\sigma_{B_1} = \sigma_{B_2} = 8.21$, $\sigma_{B_3} = 5.95$, $\sigma_{B_4} = \sigma_{B_5} = \sigma_{B_6} = 1.15$
and $\sigma_r = 0.8$\% are the uncorrelated reactor uncertainties. 
The corresponding pull parameters are ($\varepsilon_d,\eta_d,\alpha_r$).
With this choice of nuisance parameters we are able to reproduce the 1$\sigma$, 2$\sigma$ and 3$\sigma$ confidence level 
results presented in Fig.~3 of Ref. \cite{An:2013zwz} with high accuracy. 
The differences are at the level of few percent (see Tab. I and Tab. II of Ref. \cite{Girardi:2014gna}).

The T2K experiment \cite{Abe:2013hdq} consists of two separate detectors, 
both of which are 2.5 degrees off axis
of the neutrino beam. The far detector is located at 
$L_F = 295$ km from the source, the ND280 near detector 
is $L_N = 280$ metres from the target.
%%%%%%%%%%%%%%%%%%%%%%%%%%%%%%%

In our analysis we used
the public data in \cite{Abe:2013hdq,Abe:2013fuq}. 
The neutrino flux has been estimated following \cite{Abe:2013jth}.
We fixed the fiducial mass of the near and the far detector respectively as 
$FM_{\rm ND280} = 1529$ Kg and $FM_{\rm SK} = 22.5$ Kton \cite{Meloni:2012fq}; a bin
to bin normalization has been fixed in order to reproduce the T2K best fit events.
For the energy resolution function we adopt the same Gaussian form of Eq.~(\ref{Eq:res})
with $(\alpha,\beta,\gamma) = (0,0,0.085)$ GeV.

The $\chi^2_{T2K}$ is defined as:
\be
\begin{split}  \label{eqn:chispec2}
\chi^2_{T2K} (\theta,\Delta m^2, \vec S,\rho,\Omega_d,\alpha_d) & = 
\sum_{d=1}^{2}\sum_{i=1}^{n_{bins}^d}
 2 \left[M_i^d  -T_i^d \cdot \left(1 +
 \rho + \Omega_d  \right) 
  + M_i^d \log \frac{M_i^d}{T_i^d \cdot \left(1 +
 \rho + \Omega_d  \right)} \right]
  \\
 & 
 + 
\frac{\rho^2}{\sigma^2_{\rho}}
 + \sum_{d=1}^{2}
\frac{\Omega_d^2}{\sigma_{\Omega_d}^2} + {\rm Priors}
 \,. \\
\end{split}
\ee
%%%%%%%%%%%%%%%%%%%%%%%%%%%%%%%
In the previous formula, $\vec S$ is a vector containing the new physics 
parameters, $M_i^d$ are the measured events,
including the backgrounds (extracted from Fig.~4 of \cite{Abe:2013hdq}), of the d-th detector
in the i-th bin, $T_i^d = T_i^d(\theta,\Delta m^2,\vec S,\alpha_d)$
are the theoretical predictions for the rates, $\theta$ and $\Delta m^2$ are respectively 
the mixing angles and the squared mass differences contained in the oscillation
probability, $n_{bins}^d$ is the number of bins for the d-th detector.
The parameter $\sigma_{\rho}$ contains the flux, the uncorrelated $\nu$ interaction and the final-state interactions uncertainties ($\sigma_{\rho} = 8.8 \%$ Tab. II of \cite{Abe:2013hdq}), $\sigma_{\Omega_d}$
the fiducial mass uncertainty for the d-th detector ($\sigma_{\Omega_d}$ has been estimated to be $\sigma_{\Omega_d} = 1 \%$ for the far
and the near detectors similarly to \cite{Huber:2003pm}), $\alpha_d$ are free parameters which represent 
the energy scale for predicted signal events with uncertainty $\sigma_{\alpha_d}$, ($\sigma_{\alpha_d} = 1 \%$ \cite{Coloma:2012ji}).

The corresponding pull parameters are ($\rho,\Omega_d, \alpha_d$). The measured event rates at the near detector have
been estimated rescaling the non oscillated measured event rates at the far detector using the scale factor $L^2_F / L^2_N \times  FM_{\rm ND280} / FM_{\rm SK}$.
Our definition of the $\chi^2$ allows to reproduce with high accuracy the 68\% and 90\% confidence level regions for 
$\sin^2 2 \theta_{13}$ as a function of the CP violation phase $\delta$ shown in Fig.~5 of Ref. \cite{Abe:2013hdq}.

We analysed the whole Daya Bay and T2K data sample using 
$\chi^2_{tot} = \chi^2_{DB} + \chi^2_{T2K}$. We considered two different statistical analysis: i) we fixed all the standard 
oscillation parameters,
ii) we fixed all the standard oscillation parameters except $\theta_{13}$ on which we imposed a gaussian
prior defined through the mean value and the 1$\sigma$ error
$\sin^2 2 \theta_{13} = 0.140 \pm 0.038$, $\sin^2 2 \theta_{13} = 0.170 \pm 0.045$
and $\sin^2 2 \theta_{13} = 0.090 \pm 0.009$, for the different cases we have analyzed.

\end{document}